\documentstyle[aps,preprint,floats]{revtex}

\newcommand{\beq}{\begin{equation}}
\newcommand{\eeq}{\end{equation}}
\newcommand{\ds}{\displaystyle}

\newcommand{\be}{\begin{equation}}
\newcommand{\en}{\end{equation}}
\newcommand{\bea}{\begin{eqnarray}}
\newcommand{\ena}{\end{eqnarray}}

\begin{document}

\baselineskip 24pt
\title{A Matter Scalar Field in a Closed Universe}
\author{Norman Cruz$^1$\footnote{E-Mail: ncruz@lauca.usach.cl},
Sergio del Campo$^2\footnote{E-Mail: sdelcamp@aix1.ucv.cl}$
and Ramon Herrera$^{1,2}$\footnote{E-Mail: rherrera@lauca.usach.cl}}
\address{$^1$ Depto. de F\'\i sica, Universidad de Santiago de Chile, \\
Casilla 307, Correo 2, Santiago, Chile.}
\address{$^2$ Instituto de F\'\i sica, Universidad Cat\'olica de Valpara\'\i so, \\
Av Brasil 2950, Valpara\'\i so, Chile.}
\date{April 1997}
\maketitle

\begin{abstract}
\baselineskip 24pt
We investigate the possibility that the matter of the universe has a
significant component (the quintessence component) determined
by the equation of state
$p=w\rho$, with $w<0$. Here, we find conditions under which
a closed model may look like a flat Friedmann-Robertson-Walker universe
at low redshift. We study this problem in Einstein's general relativity
and Brans-Dicke theories. In both cases we obtain explicit expressions
for the quintessence scalar potential $V(Q)$, and the angular size as a function
of the redshift.

\end{abstract}

\newpage\

\section{\bf Introduction}

The Friedmann-Roberson-Walker (FRW) model may describe a nonflat
Universe motivated by the observational evidences that the total matter
density of the universe is different from its critical amount.
In fact, the measured matter density of baryonic and nonbaryonic
components, is less than its critical value.
However, theoretical arguments derived from inflationary
models\cite{Gu} and from some current microwave anisotropy measurements
favor a flat universe where the total energy
density equals to the critical density.

In this respect, other forms of matter (components) are added,
which contribute to the total energy density, so it
becomes possible to fullfil the prediction of inflation. Examples of
these kinds are the cosmological constant, $\Lambda $ (or vacuum
energy density), together with the cold dark matter (CDM) component,
which forms the famous $\Lambda$CDM model, which, among others,
seems to be the model which best fits existing observational
data\cite{OsSt}. 

Recent cosmological observations, including 
those related to the relation between
the magnitude and the redshift \cite{Peetal}, 
constrain the cosmological parameters. The test
of the standard model, which includes spacetime geometry, galaxy
peculiar velocities, structure formation, and early universe
physics, favors in many of these cases a flat 
universe model with the presence of a cosmological
constant\cite{Pe}. In fact, the luminosity distance-redshift
relation (the Hubble diagram) for the IA supernova seems to
indicate that the ratio of the matter content to its critical
value, $\Omega_0$, and the cosmological constant fits best the
values $\Omega_0 = 0.25$ and $\Lambda = 0.75$. 

>From a theoretical point of view, another possibility has risen,
which was to consider a closed universe. It seems that quantum field
theory is more consistent on compact spatial surfaces
than in hyperbolic spaces\cite{WhSc}. Also, in quantum cosmology,
the "birth" of a closed universe from nothing,
is considered, which is characterized by having a vanishing total energy,
momentum, and charge \cite{Vi}.

Motivated mainly by inflationary universe models, on the one hand, and
by quantum cosmology, on the other hand, we describe in this paper
the conditions under which a closed universe model may look flat
at low redshift. This kind of situation has been considered 
in the literature \cite{KaTo}. There, a closed model was considered
together with a nonrelativistic-matter
density with $\Omega_0<1$, and the openess is obtained by adding a
matter density whose equation of state is $p\,=\,-\rho /3$.
Texture or tangled strings represent this kind of equation of state
\cite{Da}. In a universe with texture, the additional energy density
is redshifted as $a^{-2}$, where $a$ is the scale factor. Thus it
mimics a negative-curvature term in Einstein's equations. As a
result, the kinematic of the model is the same as in an open
universe in which $\Omega_0\,<\,1$. The first person who studied a
universe filled with a matter content with an equation of state
given by $p = - \rho/3$ seems to be Kolb\cite{Ko}. He found that
a closed universe may expand eternally at constant velocity
(coasting cosmology). Also, he distinguishes a model universe
with multiple images at different redshifts of the same object and
a closed universe with a radius smaller than $H_0^{-1}$, among
other interesting consequences. 

Very recently, there has been quite a lot of work including in the
CDM model a new component called the "quintessence" component,
with the effective equation of state given by $p\,=\,w \rho$ with
$-1\,<\,w\,<\,0$. This is the so called QCDM
model\cite{St}. The differences between this model and the
$\Lambda$ model are, first, the $\Lambda$ model has an equation of state
with $w=-1$, whereas the $Q$ model has a greater value and
second, the energy density associated to the $Q$ field in
general varies with time, at difference of the $\Lambda$ model.
The final, and perhaps the most important difference, is that the
$Q$ model is spatially inhomogeneous and can cluster
gravitationally, whereas the $\Lambda$ model is totally
spatially uniform. This latter difference is relevant in the
sense that the fluctuations of the $Q$ field could have an
important effect on the observed cosmic microwave background
radiation and large scale structure\cite{CaDaSt}. However, it has
been noticed that a degeneracy problem for the CMB anisotropy 
arises\cite{HuDaCaSt}, since any given $\Lambda$ model seems to
be indistinguishable from the subset of quintessence models when
CMB anisotropy power spectra are compared. However, they become
different when the $w$ parameter varies rapidly or becomes a
constant restricted to $w > - \Omega_Q /2$. 
>From the observational point of view, there have been 
attempts to restrict the value of this parameter. Astronomical
observations of a type IA supernova have indicated that
for a flat universe, the ratio of the pressure of the $Q$-component 
to its density is restricted to $w < -0.6$, and if the
model condidered is open, then $w < - 0.5$ \cite{Gaetal}.
Certainly, improvement either in the study of the CMB anysotropy
or the type IA supernova will help us to elucidate the
exact amount of the $Q$ component in the matter content of the universe.

In this paper we discuss cosmological FRW models with a $Q$ 
field in both Einstein's theory of general relativity
and Brans-Dicke (BD) theories \cite{BrDi}. We shall restrict
ourself to the case in which the $w$ parameter remains constant.
We obtain the potential $V(Q)$ associated to the $Q$ field, and
also determine the angular size as a function of the redshift. 

\section{\bf {Einstein theory}}

In this section we review the situation in which the quintessence
component of the matter density, whose equation of state $p=w\rho$, with
$w$ a constant less than zero, contributes to the effective Einstein action
which is given by
\be
\ds S\,=\,\int{d^{4}x\,\sqrt{-g}\,\left [\,\frac{1}{16\pi\,G}\,R\,
+\,\frac{1}{2}\,(\partial_{\mu}Q)^2\,-\,V(Q)\,+\,L_{M} \right ] }.
\label{s1}
\en
Here, $G$ is Newton's gravitational constant, 
$R$ the scalar curvature, $Q$ the quintessence 
scalar field with associated potential $V(Q)$, and $L_{M}$ represents
the matter constributions other than the $Q$ component. 

Considering the FRW metric for a closed universe
\be
\ds d\,{s}^{2}\,=\,\,d\,{t}^{2}\,-\,
\,a(\,{t}\,)^{2}\,d\Omega_{k=1}^2,
\en
with $d\Omega_{k=1}^2$ representing the 
spatial line element asociated to the hypersurfaces of homogeneity,
corresponding to a three sphere, and where $a(t)$ represents
the  scale factor, which together with the assumption that the
$Q$ scalar field is homogeneous,i.e., $Q=Q({t})$, we obtain 
the following Einstein field equations:
\be
\ds H\,^{2}\,=\,\frac{8\pi\,G}{3}\,
\left (\,\rho_{M}\,+\,\rho_{Q}\,\right )\,
-\,\frac{1}{a^{2}}
\label{E1}
\en
and
\be
\ds \ddot{Q}\, +\,3\,H\,\dot{Q}\,=-
\,\frac{\partial{V(Q)}}{\partial{Q}},
\label{E2}
\en
where the overdots specify derivatives respect to ${t}$,
$H\,=\,\dot{{a}}/{a}$ defines the Hubble expansion rate, $\rho_{M}$
is the average energy density of nonrelativistic matter, and $\rho_{Q}$
is the average energy density associated to the quintessence field defined by 
$\ds \rho_{Q}\,=\,\frac{1}{2}\dot{Q}^2\,+\,V(Q)\,,$
and average pressure $ \ds p_{Q}\,=\,\frac{1}{2}\dot{Q}^2\,-\,V(Q)\,$.
As was mentioned in the introduction we shall consider a model
where the Q-component has an equation of state defined by
$\,p_{Q}\,=w\rho_{Q}$, where $w$ is considered to lie in the
range $-1<w<0$, in order to be in agreement with the current
observational data\cite{CaDaSt}.  

In order to have a universe which is closed, but still have a
nonrelativistic-matter density whose value corresponds to that of
a flat universe, we should impose the following relation
\be 
\ds \rho_{Q}\,=\,\frac{3}{8\pi\,G\,a^{2}}\,.
\label{roq}
\en
This kind of situation has been recently considered in
ref.\cite{KaTo}, where a matter density with $\Omega_0<1$ in a
closed universe was described . 

Under condition (\ref{roq}), Einstein's equations becomes analogous to that
of a flat universe, in which the matter density $\rho_M$
corresponding to dust is equal to $\rho_M^0\,[a_0/a(t)]^3$,
and the scale factor $a(t)$ is given by $a_0\,(t/t_0)^{2/3}$. 
Using the expressions for $\rho_Q$ and $p_Q$ defined above, we obtain
\be
\ds Q\,(t)=\,Q_0\,\left (\frac{t}{t_0} \right )^{\frac{1}{3}}
\label{qt}
\en
with $Q_0$ defined by
$\ds Q_0=3\,\sqrt{3\,(1+w)/8\,\pi\,G}\,
(t_0/a_0))$. The quantities denoted by the subscript 0
correspond to quantities of the current epoch.

>From solution (\ref{qt}), together with the definitions of $\rho_Q$ and
$p_Q$ we obtain an expression for the scalar potential $V(Q)$ given by
\be
\ds V(Q)\,=V_0\,\left (\frac{Q_0}{Q} \right )^4\,,
\label{vq}
\en
where $V_0$ is the present value of the scalar quintessence potential
given by $\ds V_0=3\,(1-w)/16\,\pi\,G\,a_0^2$. 

When both solutions (\ref{qt}) and (\ref{vq}) are introduced into the field
equation (\ref{E2}) the $w$ parameter necessarily will be equal
to $\frac{-1}{3}$, as is expected from the approach followed in
ref\cite{KaTo}. 

To see that a closed model at low redshift is indistinguishable from a flat
one, we could consider the angular size or the number-redshift relation as a
function of the redshift $z$, as was done in Ref. \cite{KaTo}. Here, we shall
restrict ourselves to consider the angular size only. The results
will be compared with the corresponding analogous results obtained
in BD theory.

The angular-diameter distance $d_A$ between a source at a redshift $z_2$
and $z_1< z_2$, is defined by
\be
\ds
d_A(z_1,z_2)\,=\,\frac{a_0\,sin\left [\bigtriangleup \chi(z_1,z_2) \right]}
{1+z_2},
\en
where $\bigtriangleup \chi(z_1,z_2)$ is the polar-coordinate distance
between a source at $z_1$ and another at $z_2$, in the same line of sight,
(in a flat background) and is given by
\be
\ds
\bigtriangleup \chi(z_1,z_2)\,=\,\frac{2}{a_0\,H_0}\,
\left [ \frac{1}{\sqrt{(1+z_1)}}-\frac{1}{\sqrt{(1+z_2)}} \right].
\en
Here, $H_0$ corresponds to the present value of the Hubble constant,
defined by $\ds H_0\,=\,\sqrt{8\,\pi\,G\,\rho_M^0/3}$. The corresponding
angular size of an object of proper length $l$ at a redshift $z$ results
in $\Theta\,\simeq\,l/d_A(0,z)$, which becomes (in units of $l\,H_0$)
\be
\ds
\Theta\,=\,\frac{1}{a_0\,H_0}\,\ds \frac{1+z}
{sin\left \{\frac{2}{a_0\,H_0}\left [ 1-\frac{1}{\sqrt{1+z}}
 \right ] \right \}}.
\en

For a small redshift (or equivalently, for a small time interval) the
angular size is given by 
\be
\ds
\Theta\,\simeq \,\frac{1}{z}\,+\,\frac{7}{4}\,+\,
\left [\frac{6}{(a_0\,H_0)^2}+\frac{33}{48} \right ] z + O (z^2).
\label{exp1}
\en
Since for $\Omega_0 >1$ it is found that
\be
\ds
\Theta\,=\,\sqrt{\Omega\,-\,1}\,\frac{1+z}
{sin \left \{ 2\,\sqrt{\frac{\Omega-1}{\Omega_0-1}}
\left [ tan^{-1} \left ( \sqrt{\frac{\Omega_0\,z\,+\,1}{\Omega_0\,-\,1}}
\right ) \,-\,
tan^{-1}\left (\sqrt{\frac{1}{\Omega_0\,-\,1}} \right ) \right ] \right \} },
\en
where, $\Omega$ represents the sum of matter and the
quintessence contribution
to the total matter density, we obtain that, at low redshift,
$\ds \Theta(\Omega_0>1)\,\sim\,1/z$, which coincides  with the first
term of the expantion (\ref{exp1}). Therefore, it is expected that the models
with $\Omega_0=1$ and $\Omega_0>1$ become indistinguishable at a
low enough redshift. In Fig 1  we have plotted $\Theta$ as a function of
the redshift $z$ in the range $0.01 \leq z \leq 10$, for
$\Omega_0=1$ and $\Omega_0=3/2$. We have determined the
value of $a_0\,H_0$ by fixing the polar-coordinate distance at
last scattering surface given by $\bigtriangleup \chi(z_{LS})\,=\,\pi$,
with $z_{LS}\,\simeq\,1100$, as was done in Ref. \cite{KaTo}


\section{\bf {BD Theory}}

In this section we discuss the quintessence matter model in a theory where
the "gravitational constant" is considered to be a 
time-dependent quantity. The effective action associated to the generalized
BD theory \cite{NoWa} is given by
\be
\ds S\,=\,\int{d^{4}x\,\sqrt{-g}\,\left [\,\Phi\,R\,
-\,\frac{\omega_0}{\Phi}\,(\partial_{\mu}\Phi)^2\,-\,V(\Phi)\,
+\,\frac{1}{2}\,(\partial_{\mu}Q)^2\,-\,V(Q)\,+\,L_{M} \right ] }.
\label{s2}
\en
where $\Phi$ is the BD scalar field related to the effective 
(Planck mass squared) value, $\omega_0$ is the BD parameter, and $V(\Phi)$ is 
a scalar potential asociated to the BD field. As in the Einstein case, the
matter Lagrangian $L_M$ is considered to be dominated by dust,
with the equation of state $p_M=0$. We also keep the
quintessence component described by the scalar field $Q$.

When the FRW closed metric is introduced into the action (\ref{s2}),
together with the assumptions that the different scalar fields
are time-dependent quantities only, the following set of field
equations are obtained 
$$
\ds H^{2}\,+H\,\left(\frac{\dot{\Phi}}{\Phi}\right)\,
=\,\frac{\omega_0}{6}\left(\frac{\dot{\Phi}}
{\Phi}\right)^{2}+\,\frac{8\pi}{3\Phi}
\left(\rho_M\,+\rho_Q\right)-\frac{1}{a^{2}}+
\frac{V(\Phi)}{6\Phi}\,,
$$
\be
\ds \ddot{\Phi} +3\,H\,\dot{\Phi}\,
+\,\frac{\Phi^{3}}{2\omega_0+3}\,\,\frac{d}{d\Phi}\left(
\frac{V(\Phi)}{\Phi^{2}}\right)\,=\,\frac{8\pi}{2\omega_0+3}\,\,
\left [\rho_M+(1-3w)\rho_Q\right]\,,
\en
and
$$
\ds \ddot{Q}\, +\,3\,H\,\dot{Q}\,=-
\,\frac{\partial{V(Q)}}{\partial{Q}}.
\label{qddt2}
$$
As before, we have taken the $Q$-component with  equation of state
$p_Q=w\rho_Q$, where $w$ will be determined later on.

In order that the model mimics a flat universe,
we impose the following conditions:
\be
\ds \frac{8\pi}{3\Phi}\,\rho_Q\,=\frac{1}{a^{2}}\,
-\,\frac{1}{6\Phi}\,V(\Phi)\,,
\label{cond3}
\en
and
\be
\ds
\rho_Q\,=\,\frac{\Phi^{3}}
{8\,\pi\,(1-3\,w)}\,\frac{d}{d\Phi}\left(\frac{V(\Phi)}{\Phi^{2}}\right)\,.
\label{cond4}
\en
Under these restrictions, the BD field equations 
become equivalent to that of a flat universe, in which
we assume a matter content dominated by dust. 
 
It is known that the solutions of the scale
factor $a(t)$ and the JBD field $\Phi(t)$ are given by
$a(t)=a_0\,(t/t_0)^{2(1+\omega_0)/(4+3\omega_0)}$ and
$\Phi(t)=\Phi_0\,(t/t_0)^2/(4+3\omega_0)$, respectively.
These solutions together with the constrain equations (\ref{cond3}) and
(\ref{cond4}) yield to the following expresion for the quintessence matter
field,
\be
\ds
Q\,(t)=\,Q_0\,\left(\frac{t}{t_0}\right)^{(3+\omega_0)/(4+3\omega_0)}
\label{qt2}
\en
where now $Q_0$ is defined by
$\ds Q_0=\sqrt{\frac{3\,(1+w)(4+3\omega_0)^2\,(3+2\omega_0)}{(3+\omega_0)
(9w+2\omega_0)}\frac{\Phi_0}{8\pi}}\,(\frac{t_0}{a_0})$, and
where, as before,  the quantities with the subscript 0 represent
the actual values. Notice that this result reduces
to Einstein solution, (Eq. (\ref{qt})), for
$\omega_0 \longrightarrow \infty$, together with the identification of
the gravitational constant, $\Phi_0=1/G$.

Equation (\ref{qt2}) together with equations (\ref{cond3})
and (\ref{cond4}) yields the potential associated to the BD field 
\be
 V(\Phi)\,=\,\left \{ \begin{array}{ll}
   \ds V(\Phi_0)\,\left (\frac{\Phi}{\Phi_0} \right )^{9w} & \mbox{if
   $1+2\omega_0+9w=0$}, \\
   \ds V(\Phi_0)\,\left (\frac{\Phi_0}{\Phi} \right )^{1+2\omega_0} & \mbox{if
   $1+2\omega_0+9w \neq 0$}, 
   \end{array}
   \right.
\label{Po2}
\en
where $V(\Phi_0)$ is given by
\be
\ds V(\Phi_0)\,=\,\left \{ \begin{array}{ll}
   \ds 3(1-3w)\,\left (\frac{\Phi_0}{a_0^2} \right ) & \mbox{if
   $1+2\omega_0+9w=0$}, \\
   \ds -3\left (\frac{1-3w}{2\omega_0+9w} \right )\,
   \left (\frac{\Phi_0}{a_0^2} \right )^{1+2\omega_0} & \mbox{if
   $1+2\omega_0+9w \neq 0$}. 
   \end{array}
   \right. 
\en

We shall considered the second case only, i.e. $1+2\omega_0+9w \neq 0$.
The first case gives $\ds w=-\frac{1}{9}(1+2\omega_0)$, and since
$\omega_0>500$, in agreement with solar system gravity experiments,
one obtaines $w \ll -1$, which results inappropiated for describing 
the present astronomical observational data. Notice that the second case
gives a lower bound for the parameter $w$, given by
$w>-\frac{1}{9}(1+2\omega_0)$. However, the experiments motive
us to only consider the range $-1<w<0$.

>From Eq. (\ref{Po2}), together with $\ds V(Q) =
\frac{1}{2}\,[(1-w)/(1+w)] \dot{Q}^2$ we obtain 
\be
\ds
V(Q)=V(Q_0)\,(\frac{Q_0}{Q})^{2(1+2\omega_0)/(3+\omega_0)},
\label{Pot3}
\en
where $V(Q_0)$ is defined by $\ds V(Q_0)=\frac{3(1-w)(3+2\omega_0)}
{9w+2\omega_0}\,\frac{\Phi_0}{16\pi}\frac{1}{a_0^2}$.

When these solutions are plugged into the evolution of the Q-field equation
of motion, we find that the paremeter $w$ is given by
$\ds w=-\frac{1}{3}\,\left (\frac{2+\omega_0}{1+\omega_0}\right )$,
for this equation to be valid. Note also that if
$\ds w \longrightarrow -\frac{1}{3}$ in the Einstein limit,
$\omega_0 \longrightarrow \infty$.

The corresponding angular size (in units of $l H_0$)
for this kind of theory is found to be
\be
\ds
\Theta\,=\,\frac{1}{a_0\,H_0}\,\frac{1+z}
{sin\left \{\frac{2}{a_0\,H_0}\,\alpha(\omega_0)\,
\left [ 1-(1+z)^{-\beta(\omega_0)/2}
 \right ] \right \}},
\en
where, $\ds \alpha(\omega_0)\,=\,\frac{
\sqrt{\omega_0^2+\frac{17}{6}\omega_0+2}}{\omega_0+2}$, 
$\ds \beta(\omega_0)\,=\,\frac{\omega_0+2}{\omega_0+1}$ and
$\ds H_0\,=\,\sqrt{\frac{8\,\Pi\rho_M^0}{3\,\phi_0}}$. In Fig. 2
we have plotted $\Theta$ as a function of $z$ in the Einstein
theory and Brans-Dicke theory with $\omega_0 = 500$.
Note that at $z \sim 10$ or greater, they start to become different. 

Since $\omega_0 \gg 1$ and if we take only the first-order term
in $1/\omega_0$, we obtain that 
\be
\ds
\Theta^{BD}\,=\,\Theta^{E}\,+\,
\frac{1+z}{(a_0\,H_0)^2}\,\frac{cos[\frac{2\,Z(z)}{a_0\,H_0}]}
{sin^2 [ \frac{2 Z(z)}{(a_0\,H_0)^2}]}
\left \{ 2 [Z(z)+1]\,ln(Z(z)+1)-\frac{7}{6} Z(z)
\right \} \,\frac{1}{\omega_0}\,+\,O(\frac{1}{\omega_0})^2,
\en
where $Z(z)=\sqrt{1+z}-1$, and $\Theta^{BD}$ and
$\Theta^{E}$ represent the angular size for Brans-Dicke
and Einstein theories, respectively. At $z=z_{LS}\simeq 1100$ 
the difference $\ds \Delta\Theta \equiv
\Theta^{BD}-\Theta^{E} $, becomes
$\Delta\Theta \simeq 147\left (\frac{1}{\omega_0} \right )$,
which for $\omega_0 \sim 500$
becomes $\Delta\Theta\sim 0.3$. This difference for $z=1$, with
the same value of $\omega_0$, becomes $\Delta\Theta \sim 0.05$.
Thus, we observe that this difference increases as $z$
increases,i.e., as time becomes more remote the difference
between the angular size in both theories becomes stronger. 
Figure 3 shows how this difference becomes more important at a redshift closed
to last scattering values. This difference clearly is hard to be
detected  experimentally. However, they may be an indication that
both theories become very different at $z_{LS}$, and it is
probably to search for another observable that might
distinguish between these two possibilities. Perhaps, the
spectrum due to different matter components 
may be the answer. They certainly have an effect on the cosmic
background radiation, which in principle could be observable via
temperature fluctuations. 


\section{\bf {Conclusions }}

Assuming an effective erquation of state for the $Q$ field given by 
$p = w \rho$ with negative $w$, we have computed the form of the potential 
$V(Q)$ for the Q-field in the model where a closed universe
looks similar to a flat one at low redshifts. We have found it to vary as 
$V(Q) \sim Q^{-\alpha}$, where the parameter $\alpha$ becomes a function
of the BD parameter $\omega_0$. This parameter has the correct 
Einstein limit, since for $\omega_0 \longrightarrow \infty$ this 
paramater becomes equal to 4. We have also determined the angular size
(in unit of $lH_0$) as a function of the redshifts, in both theories.
Our conclusion is that the angular size at high redshift (closed
to last scattering values) could 
distinguish between Einstein and Brans-Dicke theories.

\section{\bf Acknowledgements}

SdC was supported by Comision Nacional de Ciencias y Tecnologia 
through FONDECYT Grant N$^0$. 1971157
and UCV-DGIP Grant N$^0$ 123.744/98. NC
was supported by USACH-DICYT under Grant N$^0$ 0497-31CM. 


\newpage
 
{\Large {\bf {Figure Captions}}}
\vspace{0.5cm}

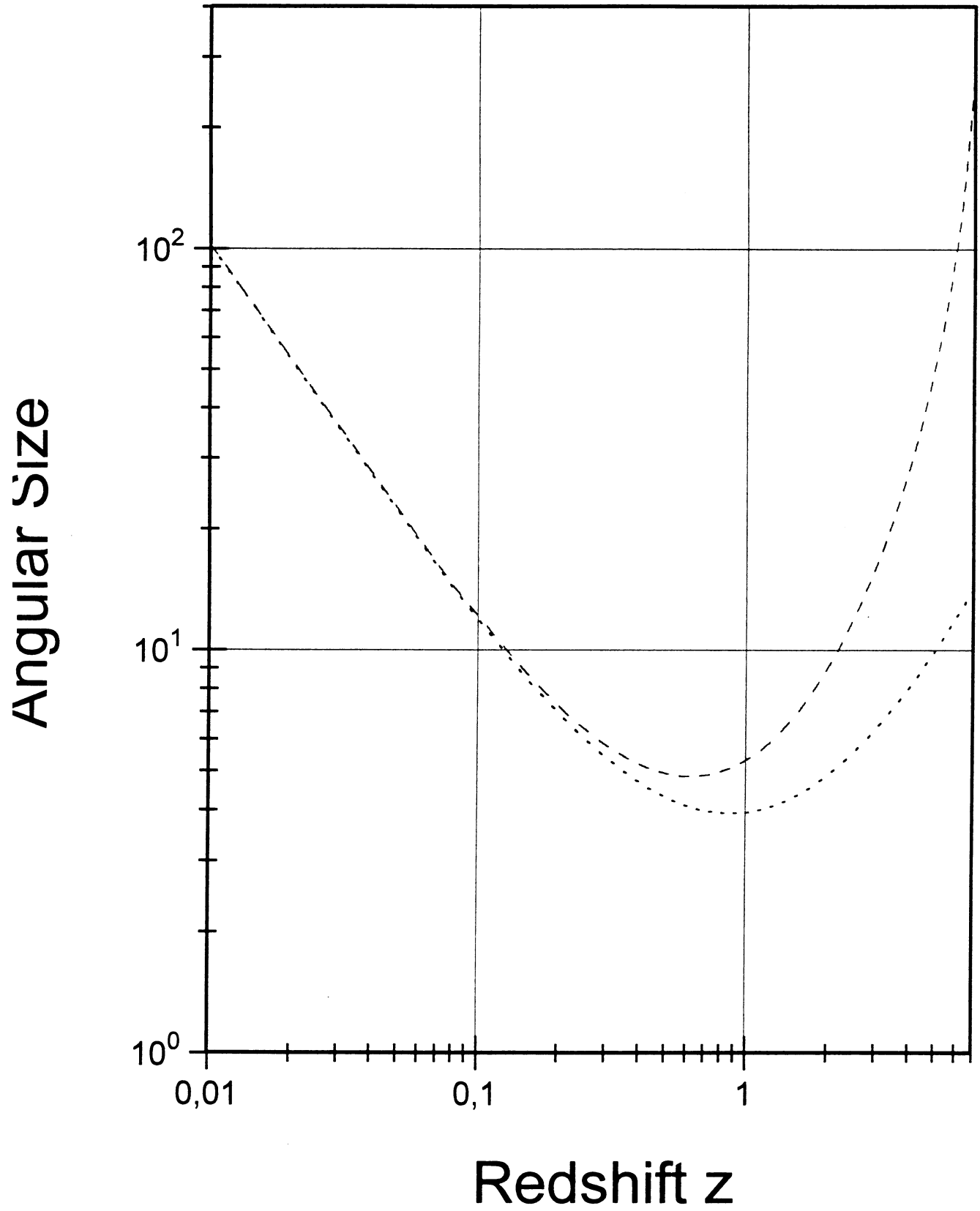
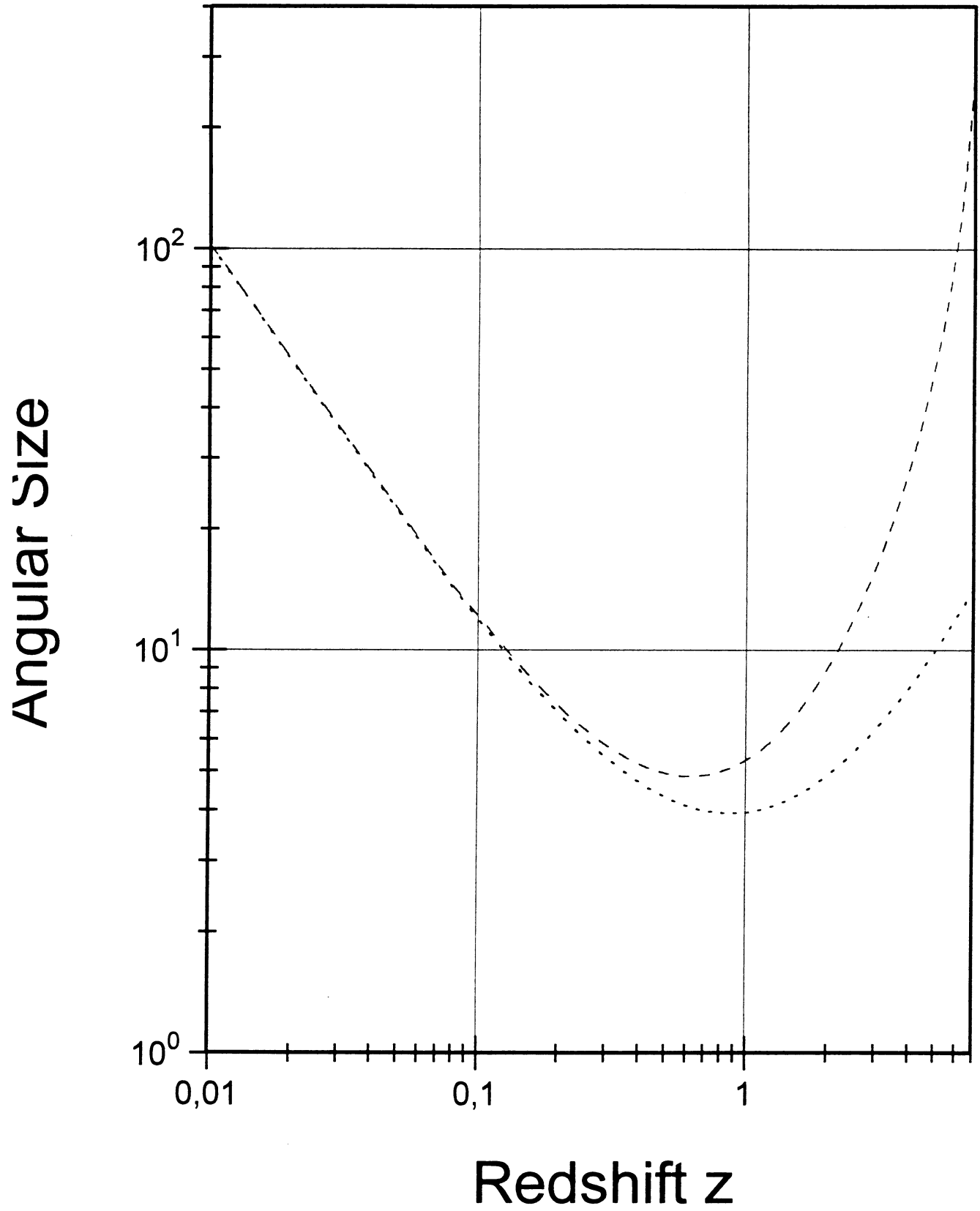
{\bf {Figure 1.-}} We plot the angular size
(in unit of $lH_0$) as a function of
the redshift, in Einstein the theory. The dotted curve corresponds to
a flat $\Omega_0=1$ universe. The solid curve represents a closed universe,
with $\Omega_0 =3/2$.

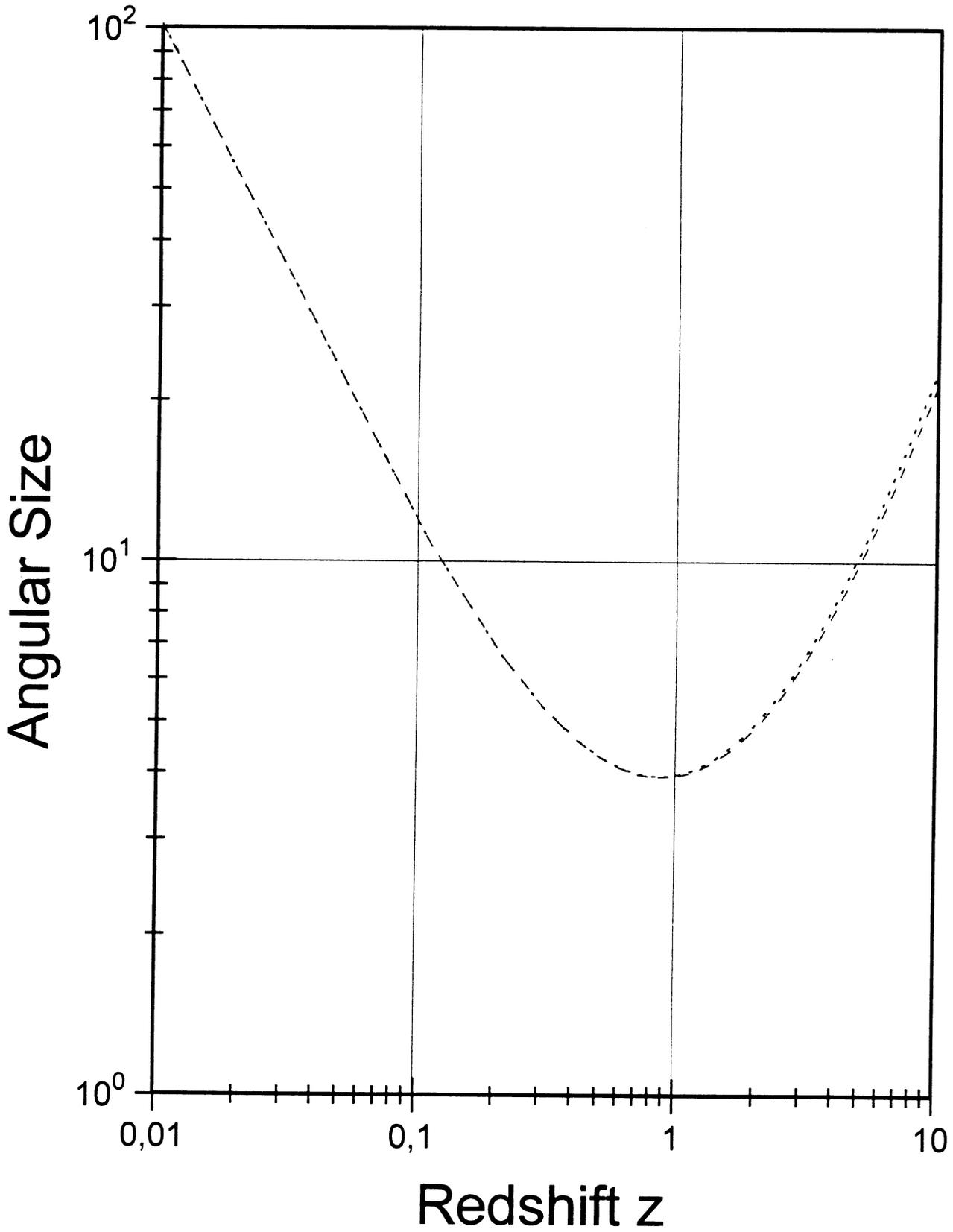
{\bf {Figure 2.-}} This plot shows how the angular size
in the Einstein (dashed
line) and Brans-Dicke (dotted curve) theories depend on the redshift for
a flat universe ($\Omega_0=1$). We have used the value $\omega_0 = 500$
for the BD parameter.

{\bf {Figure 3.-}} This plot is the same as Figure 2, but now
the range for $z$ is $10 \leq z \leq 1000$.

\end{document}